\def\year{2020}\relax
\documentclass[letterpaper]{article} % DO NOT CHANGE THIS
\usepackage{aaai20}  % DO NOT CHANGE THIS
\usepackage{times}  % DO NOT CHANGE THIS
\usepackage{helvet} % DO NOT CHANGE THIS
\usepackage{courier}  % DO NOT CHANGE THIS
\usepackage[hyphens]{url}  % DO NOT CHANGE THIS
\usepackage{graphicx} % DO NOT CHANGE THIS

\usepackage{amsmath}
\usepackage{makecell}

\usepackage{graphicx}  % DO NOT CHANGE THIS
\title{AAAI Press Formatting Instructions \\for Authors Using \LaTeX{} --- A Guide }
\author{Written by AAAI Press Staff\textsuperscript{\rm 1}\thanks{Primarily Mike Hamilton of the Live Oak Press, LLC, with help from the AAAI Publications Committee}\\ \Large \textbf{AAAI Style Contributions by
Pater Patel Schneider,} \\ \Large \textbf{Sunil Issar, J. Scott Penberthy, George Ferguson, Hans Guesgen}\\ % All authors must be in the same font size and format. Use \Large and \textbf to achieve this result when breaking a line
}

\title{Scientific Calculator for Designing Trojan Detectors in Neural Networks}

\author{
 \Large \textbf{ Peter Bajcsy\textsuperscript{\rm 1},  Nicholas J. Schaub,\textsuperscript{\rm 2}
 Michael Majurski \textsuperscript{\rm 1} } \\
  \textsuperscript{\rm 1}  Information Technology Laboratory\\
  National Institute of Standards and Technology\\
  peter.bajcsy@nist.gov  and michael.majurski@nist.gov\\
  \textsuperscript{\rm 2}   National Center for Advancing Translational Sciences (NCATS) \\
  National Institutes of Health (NIH) \\
  Axle Informatics \\
  nick.schaub@nih.gov 
  }

\begin{document}

\maketitle

\begin{abstract}
This work presents a web-based interactive neural network (NN) calculator and a NN inefficiency 
measurement that has been investigated for the purpose of detecting trojans embedded in NN models. This NN Calculator is designed on top of TensorFlow Playground with in-memory storage of data and NN graphs plus coefficients. It is ``like a scientific calculator'' with analytical, visualization, and output operations performed on training datasets and NN architectures. The prototype is aaccessible at \url{https://pages.nist.gov/nn-calculator}.
The analytical capabilities include a novel measurement of NN inefficiency using modified Kullback-Liebler (KL) divergence applied to histograms of NN model states, as well as a quantification of the sensitivity to variables related to data and NNs. Both NN Calculator and KL divergence are used to devise a trojan detector approach for a variety of trojan embeddings. Experimental results document desirable properties of the KL divergence measurement with respect to NN architectures and dataset perturbations, as well as inferences about embedded trojans.
\end{abstract}

\section{Introduction}

With the widespread use of neural networks in life-critical applications, such as self-driving cars, commercial and government agencies are concerned about the security of deployed deep learning (DL) neural networks (NNs). One example is poisoning NN models during training with datasets containing triggers (trojans) for misclassification.  A trojan is defined as a specific subset of training inputs that cause modifications of the NN weights in such a way that the NN-based classifications for inputs without and with trojans will differ. For example, a trojan can be a yellow sticky inside of a STOP sign picture \cite{Xu2019} in which case  the classifications of STOP sign and STOP sign with yellow sticky will differ. When a poisoned NN model with trojans is used for inferencing, a user will not know about the introduced misclassification by adversaries unless the specific input for inferencing is presented with the trojan. 

The motivation for this work is  to gain basic insights about trojans, their interactions with NN architectures, NN measurements that can indicate the presence of trojans, and what algorithmic approaches can be successful in detecting trojans for a variety of NN architectures under computational constraints. 

\begin{figure}[h]
 \resizebox{.5\textwidth}{!}{
\includegraphics[
  width=12cm,
  keepaspectratio,
]{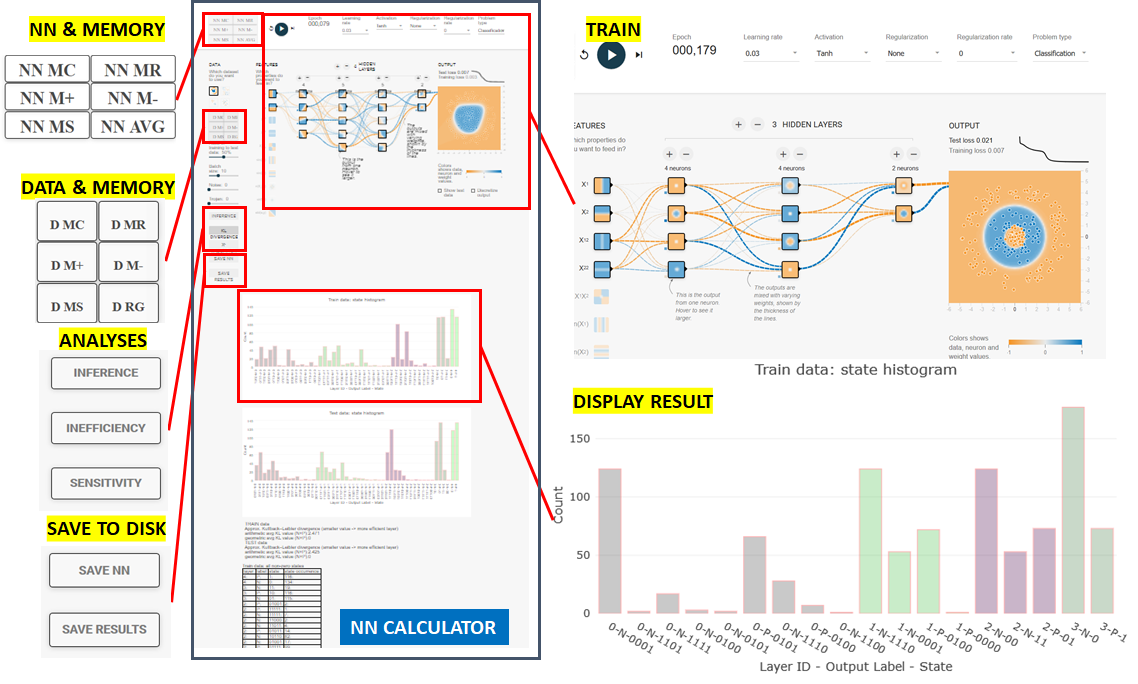}
}
  \centering
  %\fbox{\rule[-.5cm]{0cm}{4cm} \rule[-.5cm]{4cm}{0cm}}
  \caption{Interactive user interface of neural network calculator.}
%  \caption{From a scientific calculator to a neural network calculator. Arrows illustrate the extensions in terms operands, memory and NN operations, and display of results.}
  \label{fig:00}
\end{figure}

We address three problems in the aforementioned context.
The first problem is in creating an interactive environment, as shown in Figure~\ref{fig:00}, for quick evaluations of (1) NN models with varying complexities and hyper-parameters, (2) datasets with varying manifold representation complexities and class balance ratios, and (3) measurements based on varying approaches and statistical analyses. The second problem lies in designing NN efficiency measurements with understood sensitivity to variations in NN architectures, NN initialization and training, as well as dataset regeneration. The third problem is in devising an approach to detecting trojans embedded in NN models.

The problems come with associated challenges. The first challenge lies in the interactivity requirement. As of today, DL NN architectures are very complex; from 60K parameters in LeNet \cite{Khan2019}, to common networks having millions and billions of parameters (160 billion reported in \cite{Trask2015}). Modern networks require hours or days to train on advanced graphics processing unit (GPU) cards \cite{Justus2019}. 
The challenge of the second problem lies in the lack of explainable artificial intelligence (AI) \cite{Doran2018} and AI mathematical models \cite{Bruna2017}, \cite{Unser2019}, and \cite{Mallat2016}.
The last challenge lies in the large search space of possible trojans, training data, DL NN architectures, and NN training algorithms that must be understood. Related work is described in Section~\ref{section:related}.

Our approach to these challenges relies on designing a NN Calculator environment and is based on analyses of neuron states in fully connected layers of NNs. The NN Calculator is built on top of Tensorflow Playground \cite{Smilkov2017} by enabling all calculator operators on datasets and NNs, such as storing, retrieving, setting, adding, subtracting, and clearing memory containing training/testing data points and NN coefficients. Furthermore, the NN Calculator contains functionality for introducing a wide range of trojans, collecting NN state measurements, visualizing them, computing trojan sensitive probes, evaluating their robustness to NN training, and saving them for further analyses. The trade-off for interactivity of analyses is the input limitation to 2D dot patterns, the NN limitation to less than 7 hidden layers and 9 nodes per layer due to screen size, and the limitation to custom designed features derived from 2D dot patterns.

The novelty of the work lies in designing:
\begin{itemize}
	\item a web-based NN calculator for the AI community interested in gaining research insights about NN performance under various configurations,
	\item a Kullback-Liebler (KL) divergence based measurement of NN inefficiency,  
	\item an approach to detecting embedded trojans in AI models.
\end{itemize}

\section{Related Work}
\label{section:related}

The problem of detecting trojans in NN models has been posed as the Trojan in Artificial Intelligence (TrojAI) challenge by the Intelligence Advanced Research Projects Agency (IARPA) \cite{IARPA2020}. The challenges include round 0, 1, and 2 datasets consisting of trained NN models that classify input images into 5 to 25 classes of traffic signs. The goal of the challenge is to detect models trained without trojan (\texttt{TwoT}) and trained with trojan (\texttt{TwT}) based on the analyses of NN models in limited amount of time on the NIST computational infrastructure.  The problem has many variations based on what information and computational resources are available for trojan detection (type of attack, type of model architecture, model coefficients, training data subsets, description of trojans, number of classes to be misclassified by embedding trojans, classes that are misclassified by trojans, models that have been trained with trojans, computational complexity limits imposed on the delivered solution, etc.). Other challenges related to TrojAI can be found, for example, in the Guaranteeing AI Robustness against Deception (GARD) challenge \cite{Siegelmann2019}. As of today, none of the challenges can be described in terms of their difficulty level which motivates our work. 

The TrojAI challenge models were created with a variety of contiguous regions within a traffic sign defining a trojan. In the previous work, the problem of trojans in AI has been reported from the view point of detecting trojans \cite{Xu2019} \cite{Roth2019}, constructing trojan attacks \cite{Liu2018}, defending against trojans \cite{Liu2018a}, and bypassing trojan detectors \cite{Juin2019}. The problem of trojan presence is often related to the efficiency (or utilization) of DL NNs as introduced in the early publications about optimal brain \cite{LeCun1989} and optimal brain surgeon \cite{BabakHassibi1992}. A few decades later, the topics of pruning links and trimming neurons are being explored in \cite{Hu2016}, \cite{Li2017}, and \cite{Han2015} to increase an efficiency of Deep Learning (DL) NNs and to decrease NN model storage and computational requirements of model training. Our work is motivated by the past concepts of NN efficiency. However, our goal is to explore the hypothesis that NN models trained with trojans will demonstrate higher efficiency/utilization of NN than NN models trained without trojan. In comparison to previous work, our approach is focused on reliable measurements in the context of trojan detection and is investigating questions about where trojans are encoded. We assume that the models \texttt{TwoT} and \texttt{TwT} are neither under-fitted nor over-fitted \cite{Belkin2019}. 

The problem of gaining insights about DL NNs has been approached by (1) mathematical modeling \cite{Bruna2017} (network layers), \cite{Unser2019} (activation functions), \cite{Mallat2016} (wavelets), (2) feature and network visualizations  \cite{Zeiler2013} (across layers), \cite{Erhan2009}(higher layers),  \cite{Zhou2015} (discriminative features),\cite{Smilkov2017} (fully connected layers at small scale), and (3)  limited numerical precision of modeling to achieve `interactive' response \cite{Wu2016}(quantized NN for mobile devices), \cite{Rastegari2016} (binary weights for ImageNet), \cite{Gupta2015} (tradeoffs), \cite{Hubara2016} (binary NNs). Many insights are pursued with respect to representation learning \cite{Bengio2013}, expressiveness \cite{Simonyan2015a}, \cite{Lu2017}, and sensitivity and generalization (under- and over-fitting NN models) \cite{Novak2018}, \cite{Shwartz-Ziv2019}. From all past work, we leveraged the mathematical framework in \cite{Bruna2017}, visualization called Tensorflow Playground in \cite{Smilkov2017}, and efficiency and expressiveness concepts in \cite{Lu2017}. 

\section{Methods}

We describe next the developed NN Calculator with trojan simulations followed by the design of NN inefficiency measurements and our approach to trojan detection. 
 
\subsection{NN Calculator}

Our approach to designing NN Calculator aims at making it as similar as possible to a scientific calculator. 
Unlike a scientific calculator, NN Calculator operates on datasets and NN coefficients as opposed to simple numbers. Thus, we reused the symbols for $MC$, $MR$, $M+$, $M-$, and $MS$  for clearing, retrieving, adding, subtracting, and setting memory with datasets (training and testing sets) and NN coefficients (biases and weights). The user interface is shown in Figure~\ref{fig:00} (top left and middle left) where the standard five symbols are preceded with NN or D to indicate whether the operation is applied to NN or data. In addition, we included NN model averaging and dataset regeneration in order to study variability over multiple training sessions and random data perturbations. Evaluating combinations of datasets and NNs in real time enables one to explore full factorial experiments for provided factors.

%\begin{figure}
% \resizebox{.5\textwidth}{!}{
%\includegraphics[
%  width=12cm,
%  %height=8cm,
%  keepaspectratio,
%]{./figs/nncalculator2.png}
%}
%  \centering
%  %\fbox{\rule[-.5cm]{0cm}{4cm} \rule[-.5cm]{4cm}{0cm}}
%  \caption{NN Calculator user interface.}
%  \label{fig:01}
%\end{figure}

Most of the calculator settings are used for the main operations on datasets and NNs: training, inferencing, inefficiency computations, and robustness measurements (mean squared error (MSE)) for training, testing and inferencing of sub-sets. Additional operations include collecting neuron state histograms, and derived measurement statistics. 
The remaining settings are used to view characteristics of datasets (noise, trojan), parameters of NN modeling algorithm (Learning Rate, Activation Function, Regularization, Regularization Rate), and parameters of NN training algorithms (Train to Test Ratio, Batch Size). In order to keep track of all settings, we added the option of saving all NN parameters and NN coefficients, as well as saving all inefficiency and robustness analytical results. The save options are shown in Figure~\ref{fig:00} (bottom left).

\subsection{Trojan Characteristics Modeled in NN Calculator}

\begin{figure}
 \resizebox{.5\textwidth}{!}{
\includegraphics[
  width=12cm,
  keepaspectratio,
]{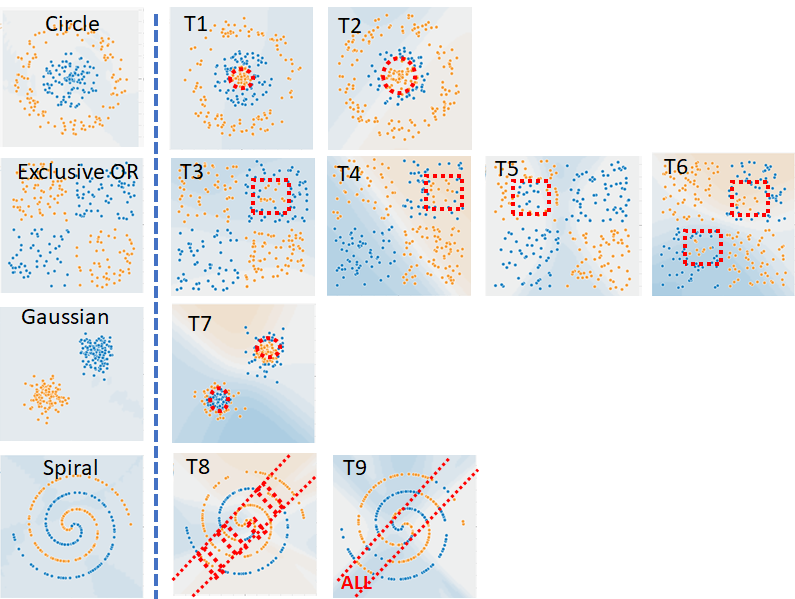}
}
  \centering
  \caption{Illustration of nine trojan embeddings in four datasets. Orange dot - class 1, blue dot - class 2, red boundary encloses dots that represent a trojan embedding.}
  \label{fig:02}
\end{figure}

In order to explore how to discriminate a model trained with trojan and a model trained without trojan, we added nine types of trojans to the NN Calculator. Our objective is to understand how the characteristics of trojans affect the trojan detection, i.e. the discrimination of models trained without trojan (\texttt{TwoT}) and trained with trojan (\texttt{TwT}). 
We generalized trojan embedding characteristics to be described by 
(1) number of trojans per class, 
(2) number of trojans per contiguous region, 
(3) shape, 
(4) size , and 
(5) location of trojans inside of a class region. 
Figure~\ref{fig:02} illustrate the nine trojan embeddings. 

\subsection{Neural Network Inefficiency Measurement}

For a given NN, its (in)efficiency is understood as the ratio of utilized representation states over the total number of representation states. Representation states are introduced next. In addition, we describe  a NN inefficiency measurement from a histogram of NN states at each layer by using (1) KL divergence, (2) a reference state distribution, and (3) computational constraints.

\begin{figure}
 \resizebox{.5\textwidth}{!}{
\includegraphics[
  width=12cm,
  keepaspectratio,
]{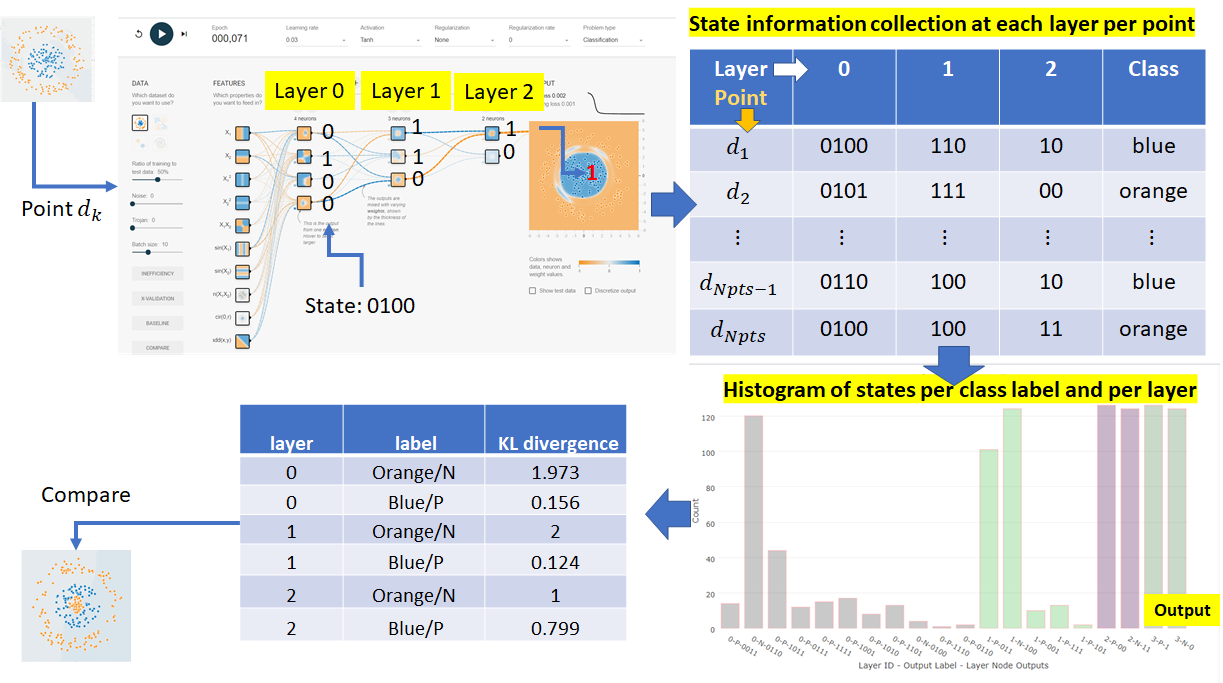}
}
  \centering
  \caption{The computation of KL divergence from NN state information at each layer per class label.  }
  \label{fig:03}
\end{figure}

\underline{States of Neural Network:} 
In order to derive NN inefficiency, we must measure and analyze states of NN layers as training data are encoded into class labels in a typical classification problem. A state of one NN layer is defined as a set of outputs from all nodes in a layer as a training data point passes through the layer.  
The output of a node is encoded as 1 if the value is positive and 0 otherwise.  
Thus, for a point $d_{k}$ from a 2D dataset with points $[d_{k}=(x_{k}, y_{k}), c_{j}]$, 
 $k=1, ..., npts$ and  $C=2$ classes $c_{1}={orange/N(negative)}, c_{2}={blue/P(positive)}$, it can generate one of  $2^{nl}$ possible states at a NN layer with $nl$ nodes. Figure \ref{fig:03} (top left) shows how a training point $d_{k}$ is converted into a feature vector that enters a neuron of the layer 0. The neuron output is generated and converted to 0 or 1 via thresholding. The neuron outputs create states 0100, 110 and 10 at the three layers for an input point. Figure \ref{fig:03} (top right) presents a table with the state information for all training points at all layers. The combined histogram of states for all  layers and both class labels (one color per layer) is shown in Figure \ref{fig:03} (bottom right).  Finally, Figure \ref{fig:03} (bottom left)  summarizes KL divergence values computed per layer and per label from the histogram of states.
 
\underline{Representation Power:} 
We view the histogram of states as a probability distribution that indicates the utilization of a layer. In order to quantify the NN utilization, we leveraged the parallels between neural network and communication fields in terms of (a) NN representation power/capacity (channel capacity in communications), (b) NN efficiency (channel efficiency), and (c) the universal approximation theorem \cite{Hornik1991} (source coding theorem \cite{Shannon1948}).
According to the universal approximation theorem, we view the NN representation power (also denoted as expressiveness or model capacity or model complexity) as its ability to assign a training class label to each training point and create accurate class regions for that class. For instance, a NN must have at least two nodes ($nl=2$) in the final layer in order to assign four class labels (i.e.,  $ C = 4 \leq 2^{nl} = 4 \rightarrow \{00, 01, 10, 11\}$).

Once we gather the state information (see Figure \ref{fig:03} (top)), we can categorize the states into three categories: 

\begin{samepage}
\begin{enumerate}
	\item State is used for predicting multiple class labels. 
	\item State is used for predicting one class label.
	\item State is not used. 
\end{enumerate}
\end{samepage}

The first category is detected when a NN does not have enough nodes (insufficient representation power). It could also occur when a NN layer does not contribute to discriminating class labels (poorly trained NN). The second category suggests that a subset of data points associated with the same class label is represented by one state (efficient or inefficient representation). 
The last category implies that a NN has a redundant (inefficient) node in a layer for representing a class label.
Thus, states at NN layers provide information about NN representation power as 
(1) \emph{insufficient,} (2) \emph{sufficient and efficient,} or (3) \emph{sufficient and inefficient.}
 An ideal NN is sufficient and efficient.
 
\underline{Inefficiency of Neural Network:} 
Since the source coding theorem is based on calculating mutual information defined via KL divergence \cite{Kullback2017}, 
we adopt KL divergence as a measurement of how inefficient it would be on average to code one histogram of NN layer states using a reference histogram as the true distribution for coding, where the reference histogram is defined below as the outcome of a uniform distribution over states assigned to each label.
Figure \ref{fig:03} (bottom) shows example results of KL divergence values derived per layer and per label that can be used to compare against values obtained from other datasets; for instance, datasets with trojans.

The rationale behind choosing entropy-based KL divergence with probability ratios comes from three considerations. First, entropy-based  measurement is appropriate because which state is assigned to predicting each class label is a random variable and a set of states assigned to predicting each class label is random. Second, probability-based measurement is needed because training data represent samples from the underlying phenomena. Furthermore,  while training data might be imbalanced (a number of samples per class varies), all training class labels are equally important and the probabilities of classes should be included in the measurement. Third, the divergence measurement reflects the fact that we measure NN efficiency relative to a maximum efficiency of NN that is achieved when sets of states utilize the entire network capacity (representation power). 

 \emph{Mathematical definition:} 
 Formally, let us denote $Q_{j}=\{ q_{ij} \}_{i=1}^{n}$ to be a discrete probability distribution function (PDF) of $n$ measured NN states and $P_{j} = \{ p_{ij} \}_{i=1}^{n}$ to be the PDF of reference (ideal) NN states. The probabilities are associated with each state (index $i$) and each class label (index $j$). The KL divergence per class label $j$ is defined at each NN layer in Equation~\ref{eq:01}.

\begin{equation*}
D_{KL}(Q_{j} \parallel P_{j})=\sum_{i=1}^{n}(q_{ij}*\log_{2} {\frac{q_{ij}}{p_{ij}}})
\tag{1}
\label{eq:01}
\end{equation*}

where $q_{ij}=\frac{count(i,j)}{p_{j}*npts}$ is the measured count of states normalized by the probability $p_{j}$ of a class label $j$ and the number of training points $npts$. 
The PDF of reference states per class label uniformly utilizes the number of states assigned to predicting each class label (i.e., 2 classes imply $\frac{1}{2}$ of all states per label). The reference probability distribution is uniform across all assigned states. Thus, all reference probabilities can be computed as $p_{ij}=m*\frac{1}{n}$ where $m$ is the number of classes and $n=2^{nl}$ is the maximum number of states ($nl$ is the number of nodes per layer).
Table~\ref{table:02} presents the theoretical definition of KL divergence with respect to input probabilities $q_{ij}$ and $p_{ij}$. 

\begin{table}
    \caption{Definition of KL divergence}
   \label{table:02}
    \centering
    \begin{tabular}{ | c | c | c |  }
      \hline
       \thead{$p_{ij}$ \textbackslash  $\quad q_{ij}$ }	& \thead{$q_{ij} = 0$} & 	\thead{$q_{ij} \neq 0$}  \\
       \hline
	  $p_{ij} = 0$     & 0 & 	\makecell{not defined} \\
	  \hline
	  $p_{ij} \neq 0$  & 0 & 	\makecell{defined} \\
       \hline
      \end{tabular}
\end{table}

Equation \ref{eq:01} for the Kullback–Leibler divergence is defined only if for all $x$, $p_{ij}=0$ implies $q_{ij}=0$. Whenever  $q_{ij}=0$ the contribution of the corresponding term is interpreted as zero because 
$\lim_{x \to 0} (x * \log_{2}x) = 0$. 
The case of ``not defined'' takes place when there are more non-zero states than the number of non-zero reference states  (i. e., the cardinality of two sets satisfies the equation: $|Set(q_{ij} \neq 0)| > |Set(p_{ij} \neq 0)|$). This case indicates that a NN has insufficient representation power to encode input dataset into a class label.

 \emph{Expected properties:}
 It is expected that KL divergence will satisfy a list of basic properties as datasets, features, and NN capacity vary. For example, given an input dataset and a set of features, inefficiency (KL divergence) per layer should increase for an increasing number of nodes per NN layer. In another example, given a NN capacity, inefficiency should decrease for datasets with added noise or trojans. The relative changes are expected to be larger than the KL divergence fluctuations due to data reshuffling, data regeneration from the same PDF or due to re-training the same NN (referred to as sensitivity of KL divergence). 

\underline{Computational Consideration about Inefficiency:} 
The KL divergence computation considers computational and memory complexities since it must scale with increasing numbers of class labels, nodes, and layers.  

 \emph{Memory concerns:} 
 One should create a histogram with the number of bins equal up to $2^{nl}$ per class label and per layer which can easily exceed the memory size. 
 For example, if a number of classes is $\approx 10$, a number of nodes is $\approx 100$, and a number of layers is $\approx 100$, then memory size is 
 $\approx 2^{100} * 10 * 100 \approx 10^{33}$  bytes.
  In our implementation approach, we create bins only for states that are created by the training data which leads to the worst case memory requirement scenario to be $npts * 10 * 100$ bytes. 
 
\emph{Computational concerns:}  
 One should align measured histograms per class label to identify the states uniquely encoding each class in order to avoid the ``not defined'' case of KL divergence or the case of the same state encoding multiple class labels. To eliminate the alignment computation in our implementation approach, we modify the KL divergence computation to approximate the KL divergence according to Equation \ref{eq:02}. The computation of modified KL divergence $\widehat{D_{KL}}$ requires only collecting non-zero occurring states and calculating their histogram. 

\begin{equation*}
\widehat{D_{KL}}(Q_{j} \parallel  P_{j})=\sum_{i \in Set(q_{ij} \neq 0)} (q_{ij} * \log_{2}{ q_{ij}} ) - \log_{2} \frac{m}{n}
\tag{2}
\label{eq:02}
\end{equation*}

While KL divergence satisfies $D_{KL} \leq 0$, the modified KL divergence $\widehat{D_{KL}}$ can be negative for those cases when  $|Set(q_{ij} \neq 0)| > |Set(p_{ij} \neq 0)|$. However, the negative value is lower bounded by Equation \ref{eq:03}. For negative values, the NN layer is insufficient for encoding input data to class labels.

\begin{align*}
\max_{Q_{j}} ( D_{KL}(Q_{j} \parallel  P_{j}) - \widehat{D_{KL}}(Q_{j} || P_{j}) ) = \notag\\
 - \sum_{i \in Set(q_{ij} \neq 0)} (q_{ij} * \log_{2}{ p_{ij}} ) - \log_{2} \frac{m}{n}
\tag{3}
\label{eq:03}
\end{align*}

The rationale behind modified KL divergence is that (1) the alignment is not important for sufficient efficient and inefficient models (it is primarily important for insufficient models), (2) the approximation assumes $p_{ij} \neq 0$ at all non-zero states $q_{ij} \neq 0$ which yields negative modified KL divergence values as indicators of insufficiency, and (3) the alignment is important for detecting poorly trained models which could be using the same states for predicting multiple class labels while leaving all other available states in a NN layer unused. For the last case, we assume that all models were properly trained and class labels are not assigned at random. Furthermore, the modified KL divergence addresses the problem of different within-class variations in training data which can lead to one class needing more allocated states than some other class. The modified KL divergence can be extended in the future by estimating within-class variations and assigning the number of states per class accordingly. In the following section we show how we use the modified KL convergence to detect the presence of trojans in a network.

\subsection{Approach to Trojan Detection}

Our assumptions are that (1) we have only testing datasets without trojans and (2) NN models with trojan and without trojan have the same accuracy. 
We can simulate many varying NN models, with 4 example datasets containing 2 classes, and nine types of trojans. 
The simulations assume close to $100 \, \%$ model accuracy on training data (with or without trojan). The  comparisons of modified KL divergence values are computed from \texttt{TwoT} and \texttt{TwT} models using datasets without trojans. The model \texttt{TwT} evaluated with datasets without trojans might have an accuracy less than $100 \, \%$ in simulations but the accuracy difference would be negligible in a real scenario (and the challenge models). 

The comparisons are performed at each NN layer and for each class label. The simulation execution is interactive (i.e., execution time is on the order of seconds) and follows the steps: 
(1) \emph{Select data}
(2) \emph{Train}
(3) \emph{Store model}
(4) \emph{Select other data}
(5) \emph{Restore model}
(6) \emph{Perform NN measurement.}

\begin{figure}
  \resizebox{.5\textwidth}{!}{
\includegraphics[
  width=15cm,
  height=6cm,
  keepaspectratio,
]{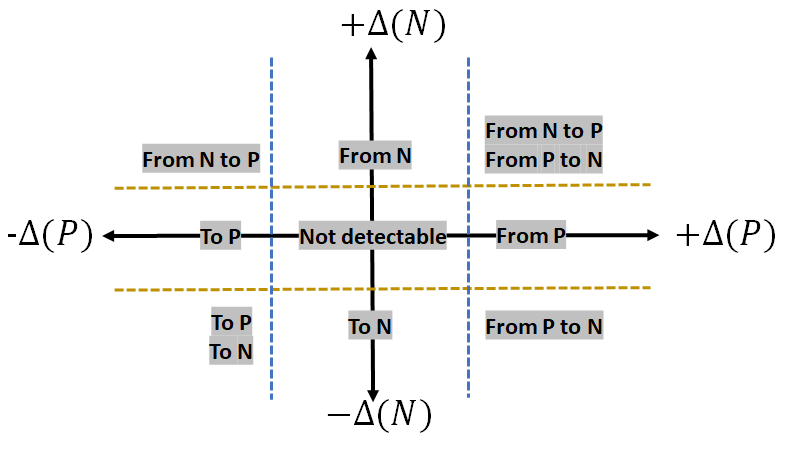}
}
  \centering
  \caption{Trojan detection using the delta between modified KL divergence of models \texttt{TwoT} and \texttt{TwT} as defined in Equation~\ref{eq:trojan}. The values for dashed lines can be determined based on the sensitivity of deltas to data regeneration and reshuffling, as well as to multiple NN initializations and re-training.
    }
  \label{fig:trojan}
\end{figure}

 Our assumption is that the magnitudes of KL divergence values for a NN model trained with a trojan embedded in a particular class (\texttt{TwT}) are smaller than the magnitudes for a NN model trained without trojan for the same class (\texttt{TwoT}). Our approach toward trojan detection is summarized in Figure~\ref{fig:trojan}. The axes correspond to the class-specific deltas between modified KL divergence of models \texttt{TwoT} and \texttt{TwT}. The dashed lines are set at a value $\sigma$ that corresponds to the sensitivity of $\widehat{ D_{KL} }$ to NN re-training as well as to data regeneration and re-shuffling. The notation ``to'' and ``from'' in Figure~\ref{fig:trojan} refers to our inference about trojans causing data points ``from'' one class to be mis-classified ``to'' another class based on the deltas defined in Equation~\ref{eq:trojan} where $P$ and $N$ are the two classes shown as blue and orange in the NN Calculator.

\begin{equation*}
	\begin{split}
\Delta(P)=\widehat{ D_{KL} }(TwoT/P) - \widehat{ D_{KL} }(TwT/P) \\
\Delta(N)=\widehat{ D_{KL} }(TwoT/N) - \widehat{ D_{KL} }(TwT/N)
	\end{split}
	\tag{4}
	\label{eq:trojan}
\end{equation*}

\section{Experimental Results}
\label{exper_results}

Next, we describe the implementation details of NN Calculator and document properties of NN inefficiency measurements.

\subsection{NN Calculator}
NN Calculator is implemented in TypeScript. The code is available from a GitHub repository with the development instructions and deployment via GitHub pages \url{https://github.com/usnistgov/nn-calculator}.
The current list of features extracted from 2D datasets includes $X1, X2, X1^2, X2^2, X1*X2, \sin(X1), \sin(X2), \sin(X1*X2), \sin(X1^2+X2^2)$, and $X1+X2$. The code uses D3.js and Plotly.js JavaScript libraries for visualization. All analytical results are displayed in NN Calculator below the NN visualization. The results consist of a state histogram (bins for both classes) and tabular summaries. The state histogram is interactive while the numerical results are presented as tables with a unique delimiter for easy parsing. 

To gain additional insights about state (although they might be computationally expensive for large NNs), simulations using NN Calculator report also the number of non-zero histogram bins per class, the states  and their counts per layer and per label for most and least frequently occurring states, the number of overlapping states across class labels and their corresponding states,  and the bits in states that are constant for all used states for predicting a class label. The additional information is reported for the purpose of exploring optimal NN architectures and investigating NN model compression schemes.

\subsection{Neural Network Inefficiency}

\underline{KL Divergence Properties:} 
We verified and quantified desirable properties of the modified KL divergence defined in Equation~\ref{eq:02}, such as decreasing inefficiency for increasing amount of added noise and increasing inefficiency for increasing number of nodes. 

\begin{figure}
 \resizebox{.5\textwidth}{!}{
\includegraphics[
  width=15cm,
  height=6cm,
  keepaspectratio,
]{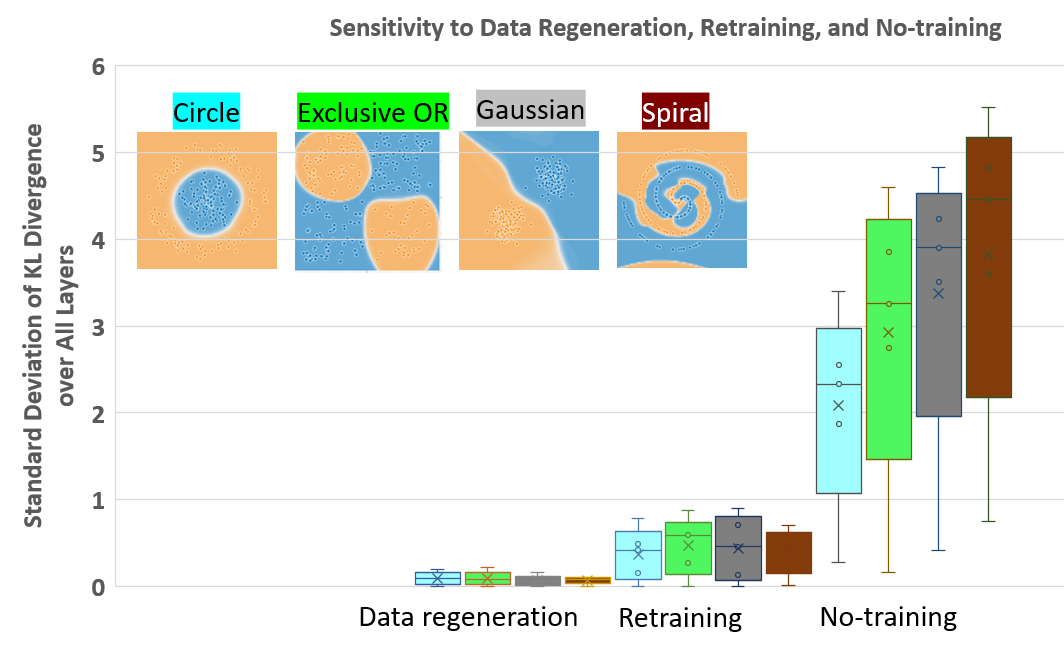}
}
  \centering
  \caption{Sensitivity of inefficiency to stochastic regeneration of datasets from the same distribution, retraining and no-training with different random initialization. The box plot shows values computed from a set of standard deviations of modified KL divergence per layer and per class for the four datasets.
 }
  \label{fig:sensit}
\end{figure}

\underline{Sensitivity of Inefficiency Measurement:} 
We quantified the sensitivity of NN inefficiency measurement with respect to (a) data reshuffling and regeneration,  (b) NN re-training with different initialization, and (c) no-training as the worst case of poor training. 
To look at the sensitivity of the NN inefficiency with respect to data regeneration, we performed the following: a NN model is trained for a dataset and stored in memory. Next, four datasets are regenerated and a standard deviation of inefficiency values are computed at each layer and for each class. Finally, the average value is computed over all standard deviations and the experiment is repeated for four 2D datasets with the results presented in Figure~\ref{fig:sensit}. From the data regeneration points in in Figure~\ref{fig:sensit},  we concluded that the average of standard deviations in inefficiency values larger than $0.1$ will indicate dissimilarity of models by other factors. 

We performed similar sensitivity experiments for no-training and retraining with random initialization. Figure \ref{fig:sensit} includes the results for four datasets. The sensitivity to retraining is bounded to approximately the average of inefficiency standard deviations equal to $0.46$ while the same value for no-training is about 5 to 8 times larger and appears to be proportional to the complexity of the class distribution.

%PB: replace with a violet plot  

\underline{Comparison of Inefficiencies for Trojan Embeddings:} 
Comparisons of models \texttt{TwoT} and \texttt{TwT} were conducted in NN Calculator using a NN with 6 hidden layers, 8 nodes per layer and 4 features including $X1, X2, X1^2, X2^2$ and $X1*X2$. The algorithmic and training parameters are set to learning rate: $0.03$, activation: $Tanh$, regularization: none, ratio of training to test data: 
$50 \, \%$, and batch size: $10$. 

\begin{figure}
 \resizebox{.5\textwidth}{!}{
\includegraphics[
  width=15cm,
  height=6cm,
  keepaspectratio,
]{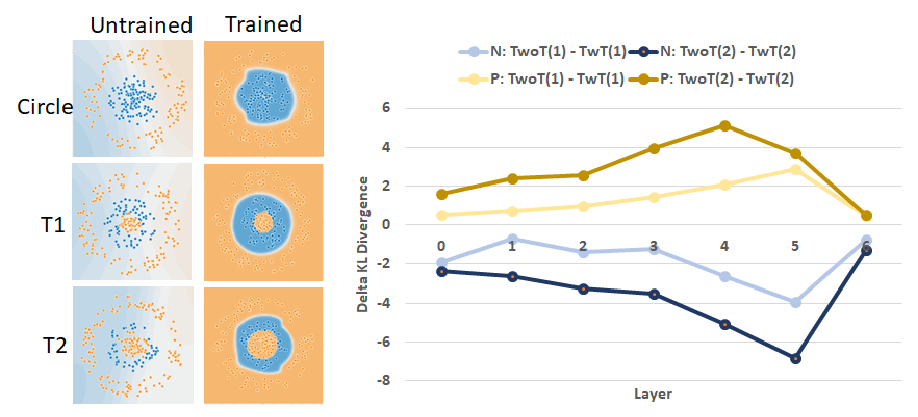}
}
  \centering
  \caption{Comparison of inefficiencies between models \texttt{TwoT} and \texttt{TwT}, and embedded orange trojans T1 and T2 with different sizes (see Figure~\ref{fig:02}, top row). The plot shows the values of $\Delta(P)$ and $\Delta(N)$ for T1 and T2 at each NN layer. 
 }
  \label{fig:08}
\end{figure}

Figure~\ref{fig:08} shows the delta between modified KL divergence values of models \texttt{TwoT} and models \texttt{TwT} for the two classes P (blue) and N (orange) and for the two trojans (T1 and T2) of different sizes (Figure~\ref{fig:08} left).  For both trojans, the delta KL divergence values are positive for the P (blue) class and negative for the N (orange) class: $\Delta(P)>0.454$ and $\Delta(N) < -0.702$. These values imply that a trojan is embedded in class P (blue) in both trojan cases and is encoding class N (orange) according to Figure~\ref{fig:trojan} (``From P to N'' $\rightarrow$ misclassified points labeled as P to N). Furthermore, as the size of a trojan increased from T1 to T2 by a size factor of 2.25, the ratio of deltas increased by $2.24$ for class N and by $2.37$ for class P.

% TODO match the orange and blue colors in the line chart to the point clouds (use two alpha values per color)

\begin{figure}
 \resizebox{.5\textwidth}{!}{
\includegraphics[
  width=15cm,
  height=6cm,
  keepaspectratio,
]{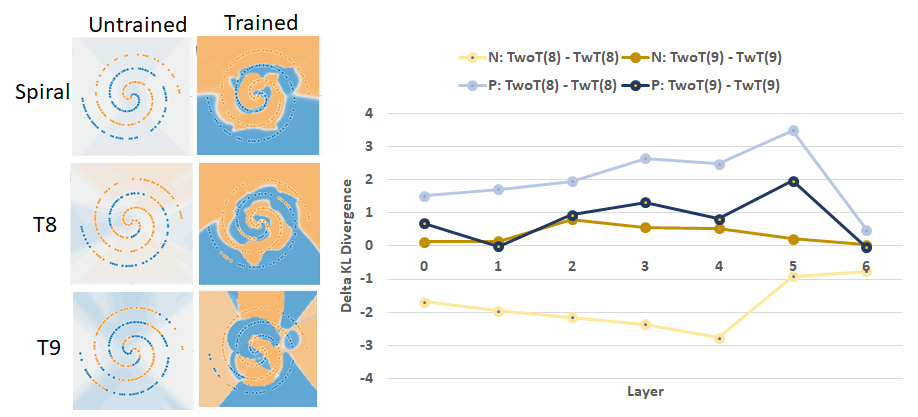}
}
  \centering
  \caption{Comparison of inefficiencies between models \texttt{TwoT} and \texttt{TwT}, and embedded trojans T8 and T9 with different number of classes (1 or 2) and class regions (1 or 4).
 }
  \label{fig:10}
\end{figure}

Figure~\ref{fig:10} illustrates the delta between modified KL divergence values of models \texttt{TwoT} and models \texttt{TwT} for the trojans T8 and T9 whose embeddings differ in terms of the number of classes and the number of class regions. First, we observe for trojan T8 that
$\Delta(T8/P) > 0.48$ and
$\Delta(T8/N) < -0.769$. These values imply that the trojan T8 is embedded in class P (blue) according to Following Figure~\ref{fig:trojan} (``From P to N''). 

We recorded much lower delta values for the trojan T9 than in the previous comparisons. This indicates the much higher complexity of modeling the spiral dataset than circle, exclusive OR, or Gaussian datasets and therefore lower inefficiency values measured at NN layers. Based on the sensitivity values shown in Figure~\ref{fig:sensit} ($0.1$ for data regeneration and $0.5$ for re-training), we could  infer that the trojan T9 is likely in both classes based on the placement of 
the point $[\Delta(T9/P) > -0.034, \; \Delta(T9/N) > 0.035]$ in Figure~\ref{fig:trojan} (i.e., the sub-spaces ``From N'', ``From P'', ``Not detectable'', and ``From N to P'' $+$ ``From P to N'').

Due to the discrete nature of the spiral pattern, the P class (blue) occupies a longer curve than the N class (orange). This contour length ratio
 ($P:N \approx 12.31:7.33$) can explain why 
 ($\Delta(T9/P) > \Delta(T9/N)$  for almost all layers. However, we are not able to make any inferences about the number of regions from Figure~\ref{fig:10} (right) other than that the complexity of modeling class P or N in the case of T8 is more inefficient than modeling class P and N in the case of T9 by comparing the deltas of modified KL divergence values.

\section{Discussion about Trojan Detection}
\label{section:discussion}

One can obtain several additional useful insights from interactive analyses in NN Calculator before designing a trojan detection algorithm. In many of the results, it is apparent that the encoded class information is not in one layer but spread across multiple layers. Thus, trojan detection must include comparisons of vectors of $\widehat{ D^{l}_{KL}}$ across all layers $l$. Furthermore, the encoding of the same training data in NN can have multiple solutions, especially in inefficient NN and therefore the comparison of vectors of $\widehat{ D^{l}_{KL}}$ must include again a statistical nature of such solutions. Finally, the last layers carry less information about trojans because they serve the purpose of a final decision maker which should appear fair for datasets without trojans. This could be accommodated by weighting the layer-specific vector elements. From a global algorithmic design perspective, designing an actual trojan detector must still consider the trade-offs of doing all pair-wise model comparisons versus clustering all vectors of $\widehat{ D^{l}_{KL}}$ to identify the cluster of model TwoT.  

\section*{Summary and Future Work}

We designed NN calculator and an inefficiency measurement for detecting trojans embedded in NN models. Our work is focused on measuring neural network inefficiency using KL divergence as a means to  advance mathematical and statistical modeling of neural networks. Current modeling efforts suffer currently from a steep learning curve, hardware requirements, and time delays between experimental runs. Some of these drawbacks can be overcome by the NN Calculator since it is interactively accessible using a browser at \url{https://pages.nist.gov/nn-calculator/} and performing experiments does not require specialized hardware (i.e., GPU cards) nor long waiting times. 

% TODO remove this before blind review since it tells people who the authors are 
\section*{Acknowledgement}
The funding for Bajcsy and Majurski was provided by IARPA, and for Schaub was provided by NCATS NIH.

% TODO remove this before blind review since it tells people who the authors are
\section*{Disclaimer}
Commercial products are identified in this document in order to specify the experimental procedure adequately.
Such identification is not intended to imply recommendation or endorsement by the National Institute
of Standards and Technology, nor is it intended to imply that the products identified are necessarily the best available for the purpose.

\bibliographystyle{aaai}
\bibliography{aaai_2020}

%\section*{References}

\end{document}